\documentclass[preprint,nofootinbib,amsmath,amssymb,aps]{revtex4-2}
\pdfoutput=1
\usepackage{dcolumn}
\usepackage{natbib}
\usepackage{graphicx}
\usepackage{comment}
\usepackage{braket}
\usepackage{indentfirst}
\usepackage{bm}

\def\xQ{\vec{x}_Q}
\def\xQc{\vec{x}_{\bar Q}}
\def\pQ{\vec{p}_Q}
\def\pQc{\vec{p}_{\bar Q}}
\def\PG{\vec{P}}
\def\p{\vec{p}}
\def\R{\vec{R}}
\def\r{\vec{r}}
\def\PCM{\vec P_{\rm CM}}
\def\k{\vec k}

\def\ta{t^a}
\def\tac{t^{a*}}
\def\1{1}
\def\Lka{L_{\k a}}
\def\Lkac{L^{\dagger}_{\k a}}
\def\Cka{C_{\k a}}
\def\Ckac{C_{\k a}^{\dagger}}
\def\Or{O_r}
\def\Lkar{L_{\k a}^{(r)}}
\def\Lkacr{L^{(r)\dagger}_{\k a}}

\begin{document}

\preprint{APS/123-QED}

\title{Quantum Brownian motion of a heavy quark pair in the quark-gluon plasma}

\author{Takahiro Miura}
 \email{miura@kern.phys.sci.osaka-u.ac.jp}
 \affiliation{Department of Physics, Osaka University, Toyonaka, Osaka, 560-0043, Japan}

\author{Yukinao Akamatsu}
 \email{akamatsu@kern.phys.sci.osaka-u.ac.jp}
 \affiliation{Department of Physics, Osaka University, Toyonaka, Osaka, 560-0043, Japan}

\author{Masayuki Asakawa}
 \email{yuki@phys.sci.osaka-u.ac.jp}
\affiliation{Department of Physics, Osaka University, Toyonaka, Osaka, 560-0043, Japan}

\author{Alexander Rothkopf}
 \email{alexander.rothkopf@uis.no}
\affiliation{Faculty of Science and Technology, University of Stavanger, 4036 Stavanger, Norway}

\date{\today}           

\begin{abstract}
In this paper we study the real-time evolution of heavy quarkonium in
the quark-gluon plasma (QGP) on the basis of the open quantum systems approach. 
In particular, we shed light on how quantum dissipation affects the dynamics of the relative motion of the quarkonium state over time. To this end we present a novel non-equilibrium master equation for the relative motion of quarkonium in a medium, starting from Lindblad operators derived systematically from quantum field theory. 
In order to implement the corresponding dynamics, we deploy the well established quantum state diffusion method. 
In turn we reveal how the full quantum evolution can be cast in the form of a stochastic non-linear Schr\"odinger equation. 
This for the first time provides a direct link from quantum chromodynamics (QCD) to phenomenological models based on non-linear Schr\"odinger equations.
Proof of principle simulations in one-dimension show that dissipative effects indeed allow the relative motion of the constituent quarks in a quarkonium at rest to thermalize. 
Dissipation turns out to be relevant already at early times well within
the QGP lifetime in relativistic heavy ion collisions.
\end{abstract}

\maketitle
\flushbottom

\section{Introduction}
\label{sec:intro}
Over the past decades, properties of nuclear matter in extreme conditions have been vigorously studied both experimentally and theoretically.

At modern collider facilities, such as the Relativistic Heavy Ion Collider (RHIC) and the Large Hadron Collider (LHC), heavy nuclei are collided at ultrarelativistic energy to create a multiparticle system in the collision center, endowed with an extremely high energy density. A multitude of measurements suggest that the energy densities reached are high enough that a new phase of nuclear matter, the ``quark-gluon plasma (QGP)'' is realized \cite{Jacak:2012dx,Adare:2006ns,Chatrchyan:2012lxa}.

On the theory side significant progress has been made in understanding the
thermodynamic, i.e. static properties of the QGP at the temperatures
reached in heavy ion collisions at which QCD dynamics is still
nonperturbative. Lattice QCD simulations have played a central role in
this regard shedding light on e.g. the crossover transition temperature
\cite{Bazavov:2014pvz,Borsanyi:2013bia}, the physics of static screening
in QCD \cite{Maezawa:2010vj,Borsanyi:2015yka} and the equilibrium
spectral properties of heavy quarkonium
\cite{Asakawa:2003re,Umeda:2002vr,Ding:2012sp,Aarts:2014cda,Ikeda:2016czj,Kelly:2018hsi,Kim:2018yhk}. On the other hand our understanding of the dynamical properties of QCD is much less developed due to the notorious sign problem preventing lattice QCD simulations from being performed directly in Minkowski time. Perturbation theory may provide some input on dynamical properties but is applicable only at much higher temperatures than achievable in experiments.

One way to handle dynamical evolution is to turn to effective field theories (EFTs), such as relativistic hydrodynamics for bulk matter or potential non-relativistic QCD for the bound states of heavy quarks, so-called quarkonium \cite{Brambilla:1999xf}. An EFT allows us to systematically simplify the theory description by focusing on the relevant degrees of freedom at a certain energy scale only. This opens up the possibility to use quantities, computable on the lattice (e.g. the equation of state), to implement a dynamical evolution of strongly interacting matter in a heavy ion collision. For the case of heavy-quark pairs it has become possible to define and derive the concept of an in-medium potential \cite{Laine:2006ns,Beraudo:2007ky,Brambilla:2008cx,Rothkopf:2011db,Burnier:2016mxc,Burnier:2014ssa}, which summarizes how the quarkonium interacts with the surrounding QGP. One central open question is how to use this in general complex valued potential to implement the microscopic evolution of the quarkonium system. In this paper we will provide a concrete example of how this may be achieved.

An understanding of the dynamical evolution of heavy quarkonium is a key factor in achieving insights into the properties of the hot matter created in heavy-ion collisions. The original proposal by Matsui and Satz, stating that the survival probability of quarkonia ($J/\psi$) may be used as a probe of QGP formation, points out one essential feature of deconfinement: Debye screening \cite{Matsui:1986dk}. In the deconfined QGP phase, light quarks and gluons that carry color are liberated and move about, screening the color charges of the heavy quarks. However, in addition to the screened potential force, the QGP constituents exert two other kinds of forces on quarkonia, namely the drag force and random force (kicks). The latter two are both qualitatively and quantitatively different from the force acting in the vacuum. 

Many different phenomenological approaches are currently used to describe quarkonium data.  
Some are based on Schr\"odinger equation with the in-medium complex potential \cite{Strickland:2011aa,Krouppa:2015yoa,Krouppa:2016jcl,Krouppa:2017jlg} and others based on kinetic Boltzmann equations with chemical reactions between a quarkonium and an unbound heavy quark pair in the QGP \cite{BraunMunzinger:2000px,Rapp:2008tf,Zhou:2014kka,Du:2017qkv,zhao2010charmonium,Yao:2018nmy,Yao:2018sgn}. It remains, however, an open question how to derive the dynamical evolution of heavy quarkonium systematically from QCD.

As heavy quarkonium constitutes a quantum mechanical bound state immersed into a strongly interacting medium, we are forced to deploy a genuinely quantum mechanical description of its dynamics. This question of the evolution of a small quantum system ``S'' coupled to an environment ``E'' has been studied in detail in condensed matter physics. 
In that context the open-quantum-system approach has been developed and
applied to the description of quarkonium in the QGP as well
\cite{Young:2010jq,Borghini:2011yq,Borghini:2011ms,Akamatsu:2011se,Akamatsu:2014qsa,Kajimoto:2017rel,Akamatsu:2018xim,Blaizot:2015hya,DeBoni:2017ocl,Blaizot:2017ypk,Blaizot:2018oev,Brambilla:2016wgg,Brambilla:2017zei,Brambilla:2019tpt,Katz:2015qja}. The
ultimate goal then is to eventually simulate a quarkonium as an open quantum system in the QGP and extract information on the in-medium forces from experimental data of $\Upsilon$ and $J/\psi$ measured at the LHC and RHIC \cite{Adare:2006ns,Adare:2008sh,Adare:2011yf,Adamczyk:2013tvk,Chatrchyan:2012lxa,Abelev:2014nua,Adare:2014hje,Khachatryan:2016xxp,Sirunyan:2018nsz,Abelev:2013ila,Adam:2016rdg}.

In the open quantum system approach \cite{breuer2002theory}, dynamical information of the small system ``S'' is encoded via the reduced density matrix $\rho_{\rm{S}}(t)$. The master equation for the evolution of the reduced density matrix can be derived by integrating out the environment degrees of freedom from the density matrix of the overall system. This process requires a set of approximations, whose validity rests on energy and time scale hierarchies between the system and the environment.

In this work, we introduce a novel master equation specifically for the relative motion of quarkonium, which turns out to be of ``quantum Brownian motion'' type. I.e., it exploits the fact that the system evolution is much slower than the environment. A general master equation for the quantum Brownian motion of quarkonia in the Lindblad form \cite{Lindblad:1975ef} has been first derived in \cite{Akamatsu:2014qsa}:
\begin{align}
\partial_t \rho_{\rm{S}}(t) = -i[H_{\rm{S}},\rho_{\rm{S}}(t)] + \sum_{i}\Bigr(2L_i \rho_{\rm{S}}(t) L_i^{\dagger} -L_i^{\dagger}L_i\rho_{\rm{S}}(t) -\rho_{\rm{S}}(t) L_i^{\dagger}L_i\Bigr),
\end{align}
where $H_{\rm{S}}$ denotes the Hamiltonian of the small system and $L_i$ the so called Lindblad operators.
This master equation describes a Markovian time evolution and one can prove that it preserves the positivity of the reduced density matrix.

The initial derivation of the Lindblad equation for quantum Brownian motion of quarkonia has led to heightened interest and activity in finding Lindblad type master equations for quarkonium in various different regimes of energy and time scale separation. One pertinent example is the quantum optical master equation for gluo-dissociation dynamics. It describes the process in which a singlet bound state (quarkonium) absorbs a real gluon and turns into an octet unbound state. It is an interesting application and at the same time requires careful examination of the validity of open system approach\footnote{
In particular, the standard approximation schemes in the literature, such as the gradient expansion (quantum Brownian motion) and the rotating wave approximation (quantum optics) \cite{breuer2002theory}, are not applicable to the transition from a {\it deeply bound} singlet state to {\it continuum} of octet spectra by gluon absorption.
It is expected that classical description by Boltzmann equation turns out to be applicable at time scale longer than the {\it decoherence} time scale as also discussed in the single heavy quark case \cite{Akamatsu:2015kaa}.
}. 

In this paper, we derive the explicit Lindblad equation for the relative motion of heavy quarkonium and present numerical simulations of its quantum Brownian motion using the quantum state diffusion (QSD) method. With the explicit inclusion of dissipative effects we overcome a central limitation of the previously deployed stochastic potential approach \cite{Akamatsu:2011se,Kajimoto:2017rel}, which corresponds to the lowest order gradient expansion of the full Lindblad equation. The present study follows in the footsteps of our recent paper \cite{Akamatsu:2018xim}, where we analyzed the quantum Brownian motion of a single heavy quark.
The main outcome was that quantum dissipation is important not only on long time scales such as the heavy quark equilibration but also in the early stages, if the initial heavy quark wave function is localized. The present paper extends this analysis to quarkonia.
With a heavy quark pair, a scattering between a heavy quark and a gluon interferes with that between a heavy antiquark and the gluon.
Therefore quantum Brownian motion for a heavy quark pair acquires nontrivial correlation between the pair, in addition to the potential force.
Using simulations in one spatial dimension via the QSD method we find that 
\begin{itemize}
\item The effective coupling of quarkonia to the QGP depends on the dipole size.
\item Study of the late time steady state of the reduced density matrix reveals that it is indeed consistent with a thermal Boltzmann distribution.
\item Inspection of the early stages reveals that the effects of quantum dissipation are also important there.
\end{itemize}

This paper is organized as follows. In Sec.~\ref{sec:oqs} we briefly
review the foundations of the open quantum system approach and introduce
a novel Lindblad master equation for the relative motion of
quarkonium. The explicit form of the Lindblad operators for the relative
motion is provided and its physical meanings are  discussed. 
We also discuss the quantum state diffusion approach, a method of
implementing the Lindblad master equation in terms of an ensemble of
stochastically evolving wave functions. 
In Sec.~\ref{sec:qsd} we present the numerical setups and then in
Sec.~\ref{sec:result} we present our numerical results from a
one-dimensional simulation for different medium conditions and study the
effect of dissipation. Section \ref{sec:summary} provides a summary and outlook.

\section{Lindblad equation for a quarkonium in the QGP}
\label{sec:oqs}
\subsection{Basics of open quantum system}
In the open quantum system approach, we consider a composite system which consists of a small system ``S'' and an environment ``E''.
The unitary evolution of the degrees freedom of ``S'' and ``E'' together can be described by a hermitian Hamiltonian $H_\text{tot}=H_{\rm{S}}\otimes I_{\rm{E}} + I_{\rm{S}}\otimes H_{\rm{E}} + H_{\rm int}$, which induces time translations in the total density matrix of states $\rho_\text{tot}$.

When we are interested in the dynamics of the small system alone, it is desirable to work with an effective description written solely in terms of the system degrees of freedom.
Therefore we turn to the reduced density matrix, defined by
\begin{align}
\rho_{\rm S} \equiv{\rm Tr}_{\rm E} \rho_\text{tot},
\end{align}
and its time evolution, usually referred to as the master equation,
\begin{align} 
\partial_t \rho_{\rm{S}}=\mathcal{L}\rho_{\rm{S}},
\label{eq:master}
\end{align}
which is the central object of the open quantum system analysis.
Here $\mathcal{L}$ refers to a superoperator that maps one reduced density matrix into another at different times and effectively describes how the environment couples to the system.

Positivity ($\forall{\alpha}, \bra{\alpha}\rho_{\rm{S}}\ket{\alpha}\geq 0$), hermiticity ($\rho_{\rm S}=\rho_{\rm S}^{\dagger}$) as well as unitarity (${\rm Tr}_{\rm S}\rho_{\rm S}=1$) are the basic properties that allow a physical interpretation of $\rho_{\rm S}$ as a density matrix.
It can be proven that if a Markovian master equation respects these conditions, it can be expressed in a particular form, the Lindblad form \cite{Lindblad:1975ef}
\footnote{
The Caldeira-Leggett master equation \cite{Caldeira:1982iu} is not in
the Lindblad form; nevertheless it describes quantum Brownian motion well.
The Lindblad master equation is, however, conceptually more useful. 
}:
\begin{align}
\partial_t \rho_{\rm{S}}(t) = -i[H_{\rm eff},\rho_{\rm{S}}(t)] + \sum_{i}\Bigr(2L_i \rho_{\rm{S}}(t) L_i^{\dagger} -L_i^{\dagger}L_i\rho_{\rm{S}}(t) -\rho_{\rm{S}}(t) L_i^{\dagger}L_i\Bigr).
\label{eq:lindeq}
\end{align}
Note that the Hamiltonian $H_{\rm eff}$ is not necessarily the same as the Hamiltonian $H_{\rm{S}}$ when the system is isolated.

In our study, the system ``S'' and the environment ``E'' correspond to a
quarkonium and the QGP (light quarks and gluons), respectively, and the Lindblad operators $L_i$, as explained below, are labeled by a continuous momentum variable $\vec{k}$ and discrete color $a$, i.e. $i=(\vec{k}, a)$.

\subsection{Lindblad equation for quarkonium relative motion}

We consider a system, where the energy density of the medium is high enough for the QCD coupling to be small and at the same time the temperature is still low enough, so that the heavy quark mass represents the largest energy scale $m_Q \gg T \gg \Lambda_{\rm QCD}$. In such a scenario, which may be realized in heavy-ion collisions at LHC, using well controlled expansions in $1/m_Q$ and the strong coupling $g$, the effective Hamiltonian and the Lindblad operators in a finite volume $L^3$ have been derived in the influence functional formalism in Ref.~\cite{Akamatsu:2014qsa}:
\begin{subequations}
\begin{align}
H_{\rm eff} &= \frac{\pQ^{\ 2}+\pQc^{\ 2}}{2M} +
\left[V(\xQ-\xQc)-\frac{1}{8MT}\left\{
(\pQ-\pQc), \vec \nabla D(\xQ - \xQc)
\right\}\right](\ta \otimes \tac),\\
\Lka&=\sqrt{\frac{\tilde D(\k)}{2L^3}}
\left[
e^{\frac{i\k\cdot\xQ}{2}}
\left(1-\frac{\k\cdot\pQ}{4MT}\right)
e^{\frac{i\k\cdot\xQ}{2}}(\ta\otimes \1)
-e^{\frac{i\k\cdot\xQc}{2}}
\left(1-\frac{\k\cdot\pQc}{4MT}\right)
e^{\frac{i\k\cdot\xQc}{2}}(\1\otimes\tac)
\right],
\end{align}
\label{eq:Lindblad}
\end{subequations}
where $(\xQ, \pQ) $ and $(\xQc, \pQc)$ denote the position and momentum
{\it operators} of the heavy quark and antiquark, respectively, and $t^a$ represent color matrices in the fundamental representation.
The in-medium real-time dynamics is hence governed by two quantities, the real functions $V(\vec r)$ and $D(\vec r)$ (and its Fourier transform $\tilde D(\vec k)$), which are defined by gluon two-point functions \footnote{
In comparison to \cite{Akamatsu:2014qsa}, we adopt $8T^2A=D$ for simplicity and opposite sign convention for $V(\vec r)$ and $D(\vec r)$ for later purposes.
} 
\begin{align}
\label{eq:DVGluonCorr}
\frac{2g^2}{N_c^2-1}\int_0^{\infty} dt \langle A_0^a(t,\vec r)A_0^a(0,\vec 0)\rangle
=D(\vec r) - iV(\vec r),
\end{align}
and given explicitly using the gluon self-energy in the hard-thermal loop (HTL) approximation,
\begin{align}
\label{eq:DV_HTL}
D(\vec r)= g^2 T\int\frac{d^3k}{(2\pi)^3}
\frac{\pi m_{\rm D}^2e^{i\vec k\cdot\vec r}}{k(k^2+ m_{\rm D}^2)^2}, \quad
V(\vec r)=  -\frac{g^2}{4\pi r}e^{-m_{\rm D}|\vec r|},
\end{align}
with the Debye screening mass $m_{\rm D}^2=(g^2T^2/3)(N_{\rm c}+N_{\rm f}/2)$ for QCD with $N_{\rm f}$ light flavors.
The Lindblad operator describes collisions between heavy quarks and plasma particles.
The operator $\exp[i\vec k\cdot\xQ](\ta\otimes \1)$ shifts the heavy quark momentum by $\vec k$ and rotates its color.
The operator with $\k\cdot\pQ/4MT$ describes the recoil of the heavy quark in each collision and accounts for its dissipation.
Note that the Lindblad operator without this dissipative term corresponds to the master equation of the stochastic potential model \cite{Akamatsu:2011se, Kajimoto:2017rel}.

The functions $V$ and $D$ in Eqs.~\eqref{eq:DVGluonCorr} and
\eqref{eq:DV_HTL} are closely related to the real and imaginary part of the complex in-medium potential $V_{Q\bar{Q}}$ derived from the Wilson loop at high temperatures \cite{Laine:2006ns,Beraudo:2007ky}. Within the HTL approximation we have
\begin{align}
{\rm Re}[V_{Q\bar{Q}}](\vec r)=V(\vec r), \quad {\rm Im}[V_{Q\bar{Q}}](\vec r)= D(\vec r)-D(0).
\end{align}
Equation \eqref{eq:Lindblad} thus provides an explicit microscopic answer to the question of how the real and imaginary part of the quarkonium in-medium potential govern the quarkonium real-time dynamics. While the real part drives the von-Neumann like part of the dynamics, it is the imaginary part that encodes how fluctuations induce the decorrelation of the quarkonium state from its initial state \cite{Akamatsu:2011se,Kajimoto:2017rel}, as it evolves over time. It is this decay in correlations and not the annihilation of the static quark pair what is encoded in the damping of the thermally averaged Wilson loop, described by its complex potential $V_{Q\bar{Q}}$.

Let us concentrate on the relative motion of the quarkonium and attempt to integrate out the center-of-mass motion. As we will see below, a complete decoupling of the center-of-mass momentum in general is not possible and it will affect the relative motion. Thus the procedure needs to be performed carefully.
Introducing the center-of-mass and the relative coordinates,
\begin{align}
\R = \frac{\xQ + \xQc}{2}, \quad
\r = \xQ - \xQc , \quad
\PG = \pQ + \pQc, \quad
\p = \frac{\pQ - \pQc}{2},
\end{align}
we can rewrite the effective Hamiltonian and the Lindblad operators as
\begin{subequations}
\begin{align}
H_{\rm eff} &= \frac{\PG^2}{4M} + \frac{\p^2}{M}
+V(\r)(\ta\otimes\tac)
-\frac{1}{4MT}\left\{ \p, \vec \nabla D(\r) \right\}(\ta\otimes\tac), \\
\Lka &=
\sqrt{\frac{\tilde D(\k)}{2L^3}} e^{i\k\cdot\R}\left[
1-\frac{\k}{4MT}\cdot\left(\frac{1}{2}\PG + \p\right) 
\right]e^{\frac{i\k\cdot\r}{2}}
(\ta\otimes\1) \nonumber\\
& \quad - \sqrt{\frac{\tilde D(\k)}{2L^3}}e^{i\k\cdot\R}\left[
1-\frac{\k}{4MT}\cdot\left(\frac{1}{2}\PG - \p\right) 
\right]e^{-\frac{i\k\cdot\r}{2}}
(\1\otimes\tac).
\end{align}
\end{subequations}
The Hilbert space for a quarkonium consists of a direct product of two Hilbert spaces: 
\begin{align}
\mathcal H_{\rm Q\bar Q} = \mathcal H_R\otimes\mathcal H_r,
\end{align}
where $\mathcal H_R$ is the Hilbert space for the center-of-mass coordinate and $\mathcal H_r$ is that for the relative coordinate and for the color space.
The density matrix for the latter is obtained by tracing out the center-of-mass coordinate
\begin{align}
\rho_r\equiv {\rm Tr}_R(\rho_{\rm Q\bar Q}).
\end{align}
To obtain the master equation for $\rho_r$, we need to calculate
\begin{align}
{\rm Tr}_R\left[H_{\rm eff}, \rho_{\rm Q\bar Q}\right], \quad
{\rm Tr}_R\left(\Lka \rho_{\rm Q\bar Q} \Lkac\right), \quad
{\rm Tr}_R\left(\Lkac \Lka\rho_{\rm Q\bar Q}\right), \quad
{\rm Tr}_R\left(\rho_{\rm Q\bar Q}\Lkac \Lka\right).
\end{align}
By expressing the Hamiltonian as $H_{\rm eff} = H_{\rm eff}^{(R)} \otimes 1 + 1\otimes H_{\rm eff}^{(r)}$, the first term is
\begin{align}
{\rm Tr}_R[H_{\rm eff}, \rho_{\rm Q\bar Q}] = \left[H_{\rm eff}^{(r)}, \rho_r\right].
\end{align}
Next, by writing the Lindblad operators as
\begin{align}
\Lka = e^{i\k\cdot\R} \Cka(\Or, \PG), \quad
\Lkac = \Ckac(\Or,\PG) e^{-i\k\cdot\R}, \quad
\Or = \left\{\r, \p, \ta\otimes\1, \1\otimes\tac \right\},
\end{align}
we obtain
\begin{subequations}
\begin{align}
{\rm Tr}_R\left(\Lka \rho_{\rm Q\bar Q} \Lkac\right)
&={\rm Tr}_R\left(\Cka(\Or,\PG) \rho_{\rm Q\bar Q} \Ckac(\Or,\PG)\right), \\
{\rm Tr}_R\left(\Lkac \Lka\rho_{\rm Q\bar Q}\right)
&={\rm Tr}_R\left(\Ckac(\Or,\PG) \Cka(\Or,\PG)\rho_{\rm Q\bar Q}\right),\\
{\rm Tr}_R\left(\rho_{\rm Q\bar Q}\Lkac \Lka\right)
&={\rm Tr}_R\left(\rho_{\rm Q\bar Q}\Ckac(\Or,\PG) \Cka(\Or,\PG)\right).
\end{align}
\end{subequations}
At this point it is not yet possible to express the above equations as explicit expressions in terms of $\rho_r={\rm Tr}_R(\rho_{\rm Q\bar Q})$. In order to proceed, let us make the assumption that the center-of-mass momentum is fixed $\PG=\PCM$. To be specific, we assume the reduced density matrix is
\begin{align}
\label{eq:assumption1}
\rho_{\rm Q\bar Q} = |\PCM\rangle\langle\PCM| \otimes \rho_r.
\end{align}
We then obtain
\begin{subequations}
\begin{align}
{\rm Tr}_R\left(\Lka \rho_{\rm Q\bar Q} \Lkac\right)
&=\Cka(\Or,\PCM) \rho_r \Ckac(\Or,\PCM), \\
{\rm Tr}_R\left(\Lkac \Lka\rho_{\rm Q\bar Q}\right)
&=\Ckac(\Or,\PCM) \Cka(\Or,\PCM)\rho_r,\\
{\rm Tr}_R\left(\rho_{\rm Q\bar Q}\Lkac \Lka\right)
&=\rho_r\Ckac(\Or,\PCM) \Cka(\Or,\PCM),
\end{align}
\end{subequations}
which tell us that for constant $\PCM$ the quantities $\Cka(\Or,\PCM)$ take the role of Lindblad operators for $\rho_r$.

Let us summarize the Lindblad master equation for the relative coordinates and color space of quarkonium
\begin{subequations}
\label{eq:Lindblad_rel}
\begin{align}
\label{eq:Linblad_rel_master}
\frac{d}{dt}\rho_r(t)&=-i[H_{\rm eff}^{(r)},\rho_r]
+\sum_{\k a}\left(
2\Lkar \rho_r \Lkacr - \Lkacr \Lkar \rho_r - \rho_r \Lkacr \Lkar
\right),\\
\label{eq:Lindblad_rel_H}
H_{\rm eff}^{(r)} &=\frac{\p^2}{M} +V(\r)(\ta\otimes\tac) -\frac{1}{4MT}\left\{ \p, \vec \nabla D(\r) \right\}(\ta\otimes\tac), \\
\label{eq:Lindblad_rel_L}
\Lkar &=\sqrt{\frac{\tilde D(\k)}{2L^3}}\left[
1-\frac{\k}{4MT}\cdot\left(\frac{1}{2}\PCM + \p\right) 
\right]e^{\frac{i\k\cdot\r}{2}}
(\ta\otimes\1)\nonumber \\
& \quad - \sqrt{\frac{\tilde D(\k)}{2L^3}}\left[
1-\frac{\k}{4MT}\cdot\left(\frac{1}{2}\PCM - \p\right) 
\right]e^{-\frac{i\k\cdot\r}{2}}
(\1\otimes\tac).
\end{align}
\end{subequations}
In this derivation, the center-of-mass momentum $\PCM$ is given as an
external parameter and can depend on time.
This explicit dependence provides a way to incorporate the relative motion of quarkonium traversing the QGP in the real-time description when the quarkonium velocity $\PCM/2M$ is small.
We find that this effect on the quarkonium relative motion is mild (see
Appendix \ref{sec:results_com} for details) and we fix $\PCM$ at $\vec 0$ in all of the simulations in Sec.~\ref{sec:result} .

Let us mention how the above master equation differs from previous proposals in the literature.
In a recent study \cite{DeBoni:2017ocl}, the effects of dissipation have
been considered in a quarkonium master equation; however, the author did
not derive a self-consistent Lindblad equation, and as a result additional
terms needed to be added by hand.
A Lindblad-like master equation in a weakly coupled setting has been derived in \cite{Blaizot:2017ypk} with a focus of further simplifying the dynamics using semi-classical approximations.
While our master equation relies on the weak-coupling approximation, we will avoid any further semi-classical approximations and instead implement the full quantum time evolution.

\subsection{Quantum state diffusion}
In quantum mechanics, we can either consider the time evolution of a
system based on its density matrix or go over to a mixed state description in terms of wave functions.
The change from one to the other is called stochastic unraveling of the master equation.
Since evolving the density matrix in the position basis incurs high numerical cost, it is often advantageous to carry out simulations in the language of wave functions. 

While the lowest order gradient expansion of the quarkonium Lindblad equation can be stochastically unravelled into unitary time evolution of wave functions based on a {\it linear} Schr\"odinger equation with a stochastic potential \cite{Akamatsu:2011se, Kajimoto:2017rel}, the full Lindblad equation requires a more sophisticated treatment.
As has been worked out in detail in the quantum physics community, any Lindblad master equation may be unravelled stochastically via the ``quantum state diffusion (QSD)'' method \cite{gisin1992quantum}.

In the QSD method, the density matrix $\rho_{\rm{S}}$ is obtained from the ensemble average of wave functions,
\begin{align}
\rho_{\rm{S}} = {\rm M}[\ket{\psi(t)}\bra{\psi(t)}],
\end{align}
and the wave functions evolve according to the following nonlinear stochastic Schr\"{o}dinger equation,
\begin{align}
\label{eq:qsd1}
\ket{d\psi}&\equiv\ket{\psi(t+dt)}-\ket{\psi(t)}
\notag\\
&=-iH_{\rm{eff}}\ket{\psi(t)}dt+\sum_{i}\Bigr(2\langle L^{\dagger}_i\rangle_{\psi} L_i-L^{\dagger}_i L_i-\langle L^{\dagger}_i \rangle_{\psi}\langle L_i \rangle_{\psi}
\Bigr)\ket{\psi(t)}dt
\notag\\
&\hspace{3.85cm}+\sum_i \Bigr(L_i- \langle L_i \rangle_{\psi}\Bigr)\ket{\psi(t)}d\xi_i\,,
\end{align}
where $\rm M$ represents taking the ensemble average and $d\xi_i$ is complex white noise satisfying
\begin{subequations}
\begin{align}
&{\rm M}[d\xi_i] = {\rm M}[\Re(d\xi_i)\Im(d\xi_j)]=0, \\
&{\rm M}[\Re(d\xi_i)\Re(d\xi_j)] ={\rm M}[\Im(d\xi_i)\Im(d\xi_j)]=\delta_{ij} dt.
\end{align}
\end{subequations}
The nonlinearity arises from the terms containing the expectation value of the Lindblad operator with respect to the wave function $\langle L_i \rangle_{\psi}$. The above QSD equation can be shown to be equivalent to the following nonlinear stochastic Schr\"{o}dinger equation for unnormalized wave functions
\begin{subequations}
\begin{align}
\label{eq:qsd2}
\ket{d\psi(t)}
&=-iH_{\rm{eff}}\ket{\psi(t)}dt+\sum_{i}\bigr(2\langle L^{\dagger}_i \rangle_{\psi} L_i-L^{\dagger}_i L_i\bigr)\ket{\psi(t)}dt
+\sum_i L_i \ket{\psi(t)}d\xi_i, \\
\label{eq:qsd}
\rho_{\rm{S}}(t)&\equiv {\rm M}\left[\,\frac{\ket{\psi(t)}\bra{\psi(t)}}{\langle{\psi(t)}|\psi(t)\rangle}\right],
\end{align}
\end{subequations}
which we employ in our numerical simulation.
When we calculate the occupation number of some specific state $\phi_i(\vec{x})$ of a quarkonium, we thus calculate
\begin{align}
N_i(t)\equiv\int d\vec x d\vec y \phi_i(\vec x)^* \rho_{\rm{S}}(\vec x,\vec y,t) \phi_i(\vec y)
={\rm M}\left[\frac{|\langle\phi_i|\psi(t)\rangle |^2}{\langle{\psi(t)}|\psi(t)\rangle}\right].
\end{align}

We would like to emphasize that with Eq.~\eqref{eq:qsd} the full quantum time evolution of the in-medium quarkonium system has been cast into the form of a stochastic {\it nonlinear} Schr\"odinger equation.
As our derivation originates in the influence functional treatment of the QCD path integral and deploys a well defined set of approximations, it is hence possible for the first time to provide a direct link between QCD and previous phenomenological models that introduce a non-linear Schr\"odinger equation in an ad-hoc manner.

\section{Numerical setup}
\label{sec:qsd}
Let us introduce our simulation prescription for the relative motion of a quarkonium in the QGP. 
The evolution equation for the wavefunction from the QSD method consists of two parts:
The effective Hamiltonian term, as well as additional terms related to the Lindblad operators (including the stochastic ones). They represent a stochastic integro-differential equation, whose explicit form for three dimensions is shown in Appendix \ref{sec:qsdeqform}.

For simplicity, in this study, we simulate the full dissipative dynamics in one spatial dimension and ignore the heavy quark colors. We solve the Hamiltonian term via the 4th order Runge-Kutta method and the Lindblad terms via a simple forward Euler time step. 

Our goal is to showcase how dissipative effects influence the evolution of quarkonium states. Thus we wish to compare to simulations in which these effects are absent. Within the QSD framework this can be achieved by discarding all terms (except for the kinetic energy) that are not finite in the $T/M\to 0$ limit. This leads to the following evolution equation:
\begin{subequations}
\begin{align}
d\psi(x)=&
-idt\left[-\frac{{\nabla}^2}{M} + V(x)\right]\psi(x)
-dt \Bigr[D(0) - D(x)\Bigr]\psi(x) \nonumber \\
&+\frac{2dt}{\bra{\psi}\psi\rangle}
\int dy \Bigr[D\Bigr(\frac{x-y}{2}\Bigr) - D\Bigr(\frac{x+y}{2}\Bigr)\Bigr][\psi^{\dagger}(y)\psi(y)]\psi(x)  \nonumber\\
&+\Bigr[d\xi\Bigr(\frac{x}{2}\Bigr) -d\xi\Bigr(\frac{-x}{2}\Bigr)\Bigr]\psi(x)
+\mathcal O(T/M), \\
{\rm M}[ d\xi(x) &d\xi^*(y) ] =D(x-y)dt,
\end{align}
\label{eq:qsd_intdiff}
\end{subequations}
which is equivalent to the master equation of the stochastic potential model \cite{Kajimoto:2017rel} in spite of rather different appearances. The former is nonlinear while the latter is linear in the wave function.

\begin{table}[tb]
\begin{center}
\caption{Numerical setup, parametrization of $V(x)$ and $D(x)$, and center-of-mass momentum}
\label{tab:setup}
\vspace{3mm}
\begin{tabular}{ccc|ccccc|c} \hline
$\Delta x$ & $\Delta t$ & $N_x$ & $\gamma$&$\ell_{\rm{corr}}$&$\alpha$&$m_{\rm{D}}$&$x_{\rm{c}}$& $P_{\rm CM}$  \\\hline 
$1/M$ & $0.1M(\Delta x)^2$ & 254 & $T/\pi$&$1/T$&$0.3$&$2T$&$1/M$& 0  \\ \hline
\end{tabular}
\end{center}
\end{table}

The parametrization for the functions $V(x)$ and $D(x)$ are given as follows:
\begin{align}
\label{eq:VD_parametrization}
V(x)=-\frac{\alpha}{\sqrt{x^2+x_{\rm{c}}^2}}\mathrm{e}^{-m_{\rm{D}}|x|}, \quad
D(x)=\gamma\exp(-x^2/\ell_{\rm corr}^2).
\end{align}
From perturbative calculations \eqref{eq:DV_HTL}, we have
\begin{align}
\alpha = \frac{g^2}{4\pi}, \quad
m_{\rm D} = gT\sqrt{\frac{N_c}{3} + \frac{N_f}{6}} \simeq \frac{2}{\ell_{\rm corr}}, \quad
\gamma = D(0) = \frac{g^2 T}{4\pi}, \quad
\label{eq:debyemass}
\end{align}
so that we choose $\alpha = 0.3$, $m_{\rm D} = 2\ell_{\rm corr}^{-1} = 2T$
\footnote{
$m_{\rm D}\simeq {2}/{\ell_{\rm corr}}$ in \eqref{eq:debyemass} is obtained by equating the full width half maximum of $D(r)$ in \eqref{eq:DV_HTL} and that in \eqref{eq:VD_parametrization}.
}, and $\gamma = T/\pi$.
Since the one-dimensional Coulomb force is singular at the origin, we regularize the potential by a cutoff $x_c=1/M$ corresponding to the validity of the nonrelativistic approximation $|p| < M$.
In the simulations in the next section, the temperature of the QGP is chosen to be 
$T/M=0.1$-$0.3$ or $T(t)=T_0\cdot [t_0/(t_0 + t)]^{1/3}$ with $t_0 =
0.84 \ {\rm fm}$ and $T_0 = 0.47 \ {\rm GeV}$ for the case of Bjorken expansion.
The center-of-mass momentum is set to $P_{\rm CM}=0$ as the dependence
of the quarkonium dynamics on its values is found to be small (see
Appendix \ref{sec:results_com} for the details).

To simulate quarkonium physics with this setup, we discretize space and time by $\Delta x = 1/M$ and $\Delta t = 0.1 M (\Delta x)^2$.
The spatial discretization $\Delta x$ is chosen much finer than the typical medium length scales $m_{\rm D}^{-1}\sim\ell_{\rm corr}/2\sim 1/2T$.
We also take the system volume $L=N_x \Delta x=254\Delta x$ much larger than the medium length scales $m_{\rm D}^{-1}\sim\ell_{\rm corr}/2\sim 1/2T$. 
These parameters and the lattice setup are summarized in Table \ref{tab:setup}.

Finally, let us remark on the boundary conditions used for the noise field $d\xi(x)$.
To simulate on a finite size lattice $[-L/2, L/2]$, we impose periodic boundary conditions on the wave function $\psi(x)$.
Then the boundary condition for the noise field is given by
\begin{align}
d\xi\left(-\frac{L}{4}\right) = d\xi\left(\frac{L}{4}\right),
\end{align}
requiring that the noise field $d\xi(x)$ obeys a periodicity of $L/2$.
Correspondingly, the noise correlation function $D(x-y)$ for $-L/4\leq x, y \leq L/4$
should be interpreted as $D(r_{xy})$:
\begin{align}
{\rm M}[ d\xi(x) &d\xi^*(y) ] =D(r_{xy})dt, \quad
r_{xy} \equiv \min\left\{
|x-y|, \frac{L}{2}-|x-y|
\right\}.
\end{align}
In this way, we redefine the function $D(x)$ for a finite system with an interval $-L/2\leq x\leq L/2$, which is all that is needed for solving the QSD equation.

\section{Numerical results}
\label{sec:result}
We study and simulate how quantum dissipation and center-of-mass motion of a quarkonium affect the dynamical evolution of its relative motion (see Appendix \ref{sec:results_com} for the latter). 
We first show in Sec.~\ref{sec:results_equilibration} how quantum dissipation influences the time evolution.
We numerically confirm that in the presence of quantum dissipation the quarkonium system approaches a steady state at late times that is independent of the initial conditions from which the evolution commenced. We furthermore find that the late time distribution is very close to the Boltzmann distribution, whose slope is well within $10\%$ of the input temperature of the surrounding medium. The quarkonium system thus appears to approach genuine thermal equilibrium at late times, assisted by the interplay of quantum fluctuations and dissipation.

We then show in Sec.~\ref{sec:results_tm} how the evolution depends on
temperature and heavy quark mass by comparing with simulations with
$T/M=$ 0.1 and 0.3.
By rescaling the evolution time $t$ by the heavy quark relaxation time,
we find that a similar relaxation behavior shows up in both of the cases.
Finally in Sec.~\ref{sec:results_pheno}, we present the quarkonium evolution in a time-dependent background with temperature $T(t)$, which decreases according to Bjorken expansion.
To make closer contact with experimental data, we project the quarkonium wave functions onto the eigenstates in Cornell potential.

\subsection{Equilibration with quantum dissipation}
\label{sec:results_equilibration}
We first simulate at $T/M=0.1$ and compare the two cases where the initial condition is either given by the ground state or the 1st excited state of the Hamiltonian
\begin{align}
H_{\rm Debye}=\frac{p^2}{M}-\frac{\alpha}{\sqrt{x^2+x_{\rm{c}}^2}}\mathrm{e}^{-m_{\rm{D}}|x|}\,.
\end{align}
Note that the Hamiltonian $H_{\rm Debye}$ here is different from the effective Hamiltonian $H_{\rm eff}$ in \eqref{eq:Lindblad_rel_H}.

\begin{figure}[htbp]
\begin{center}
\includegraphics[scale=1.0]{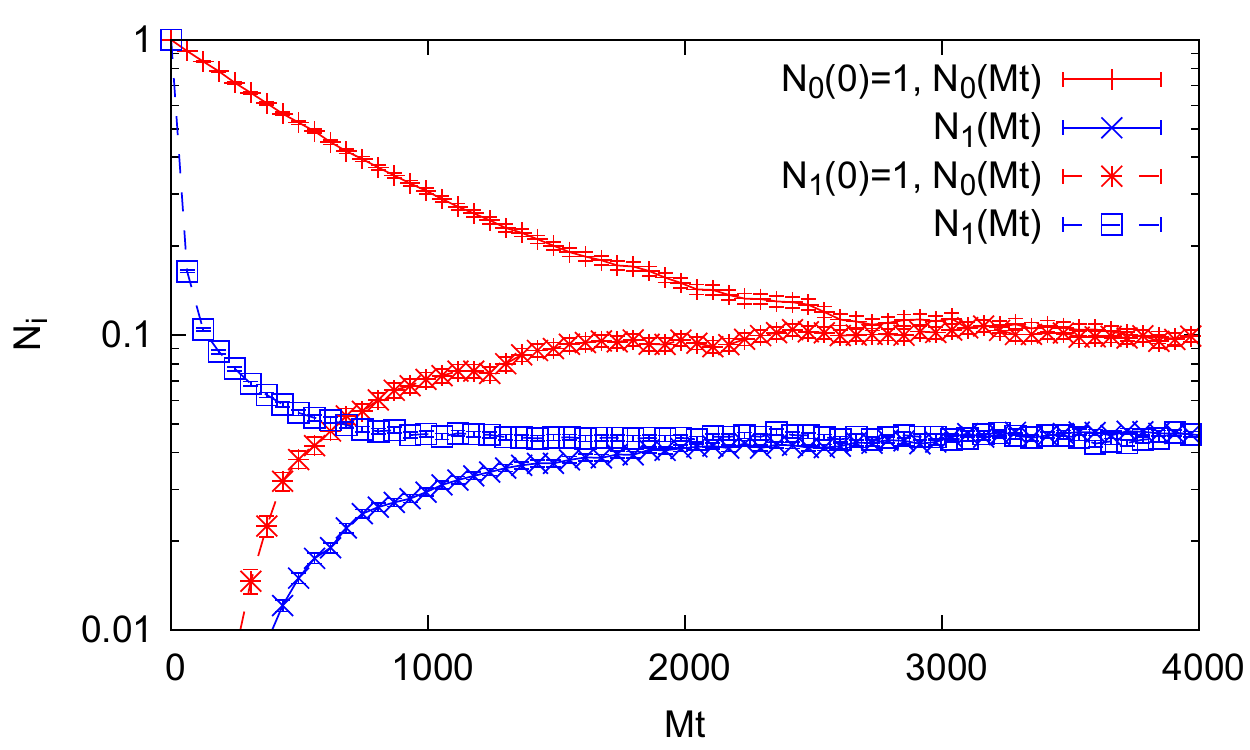}
\end{center}
\caption{
Time evolution of the occupation numbers of the ground state and the 1st excited state.
The system reaches a thermal steady state independent of the initial conditions at late times. Error bars represent the statistical errors within the ensemble average.
}
\label{fig:main_DEBYE_diss}
\end{figure}

In Fig.~\ref{fig:main_DEBYE_diss}, we plot the time evolution of the occupation number $N_i(t)$ of the $i$-th eigenstate for the ground state ($i=0$) and the 1st ($i=1$) excited state of $H_{\rm Debye}$ for two different initial conditions $N_0(0)=1$ and $N_1(0)=1$.
We find that each of the occupation numbers relaxes to a constant value irrespective of the initial condition and hence confirm that the dissipative effects lead to a steady state. This behavior is indicative that the relative motion of a quarkonium in the QGP becomes equilibrated.

\begin{figure}[htbp]
\begin{center}
\includegraphics[scale=1.0]{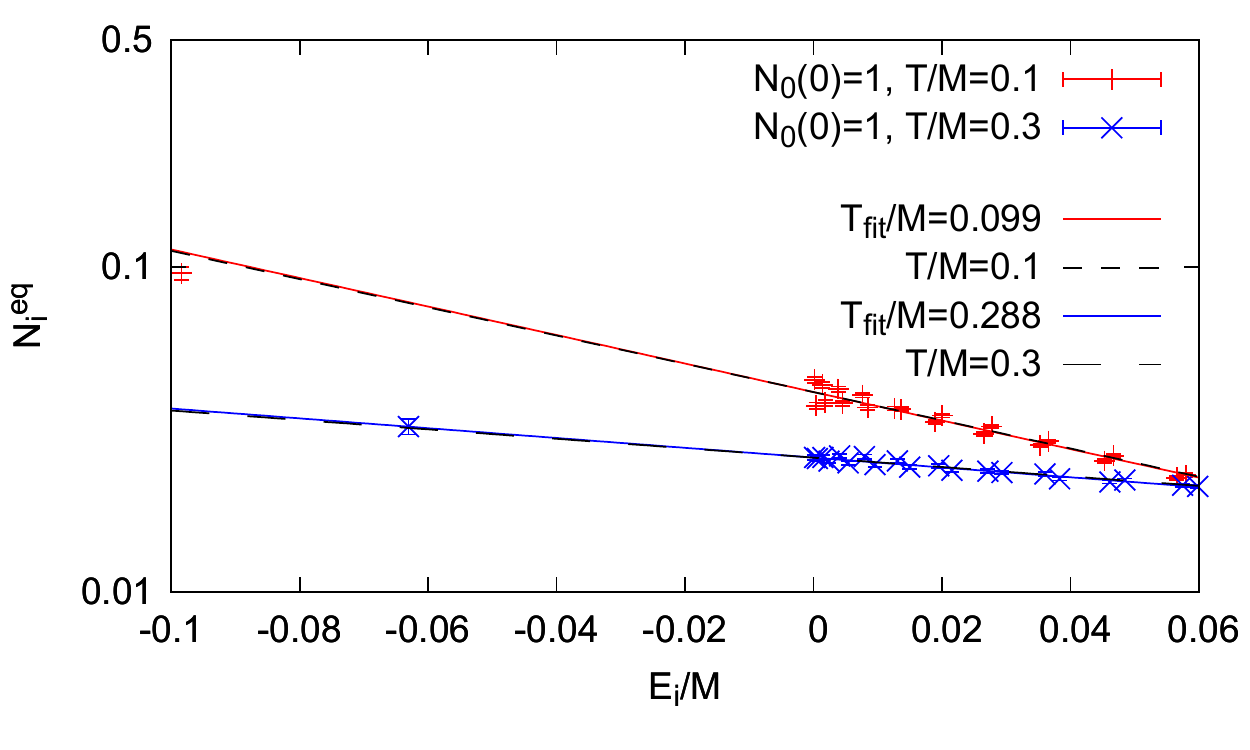}
\end{center}
\caption{
Steady state distributions of the eigenstates in $-0.10\leq E_i/M \leq 0.06$ are measured at $Mt=4650$ for $T/M=0.1$ and at $Mt=900$ for $T/M=0.3$.
The data is fitted by the Boltzmann distribution $\propto\exp[-E_i/T_{\rm fit}]$ for levels with relative velocity less than 0.5 (lowest 21 levels) for $T/M=0.1$ and $0.3$.
The fitted temperatures are $T_{\rm{fit}}/M=0.099\pm0.004$ for $T/M=0.1$ and $T_{\rm{fit}}/M=0.288\pm0.013$ for $T/M=0.3$.
}
\label{fig:main_DEBYE_temperature}
\end{figure}

Next we plot in Fig.~\ref{fig:main_DEBYE_temperature} the occupation numbers of the lower levels at late times $Mt=4650$, well within the steady state regime, as a function of the eigenenergy $E_i$ of the Hamiltonian $H_{\rm Debye}$.
We also show the results for the same analysis with $T/M=0.3$ at time $Mt=900$.
The distribution can be fitted by the Boltzmann distribution $\propto\exp(-E_i/T_{\rm{fit}})$ with $T_{\rm fit}/M = 0.099 \pm 0.004$ for $T/M=0.1$ and $T_{\rm{fit}}/M=0.288\pm 0.013$ for $T/M=0.3$.
 The fitting range is limited to the eigenstates $\phi_i$ with velocity $\frac{\sqrt{\langle p^2 \rangle_{\phi_i}}}{M/2} < 0.5$.
\footnote{
We expect that the steady state of the Lindblad evolution is the Boltzmann distribution only when the velocity is small.
See Appendix \ref{sec:eqdist_classical} for more details.
}
This corresponds to fitting the lowest 21 levels for both $T/M=0.1$ and $0.3$.

\begin{figure}[htbp]
\begin{center}
\includegraphics[scale=1.0]{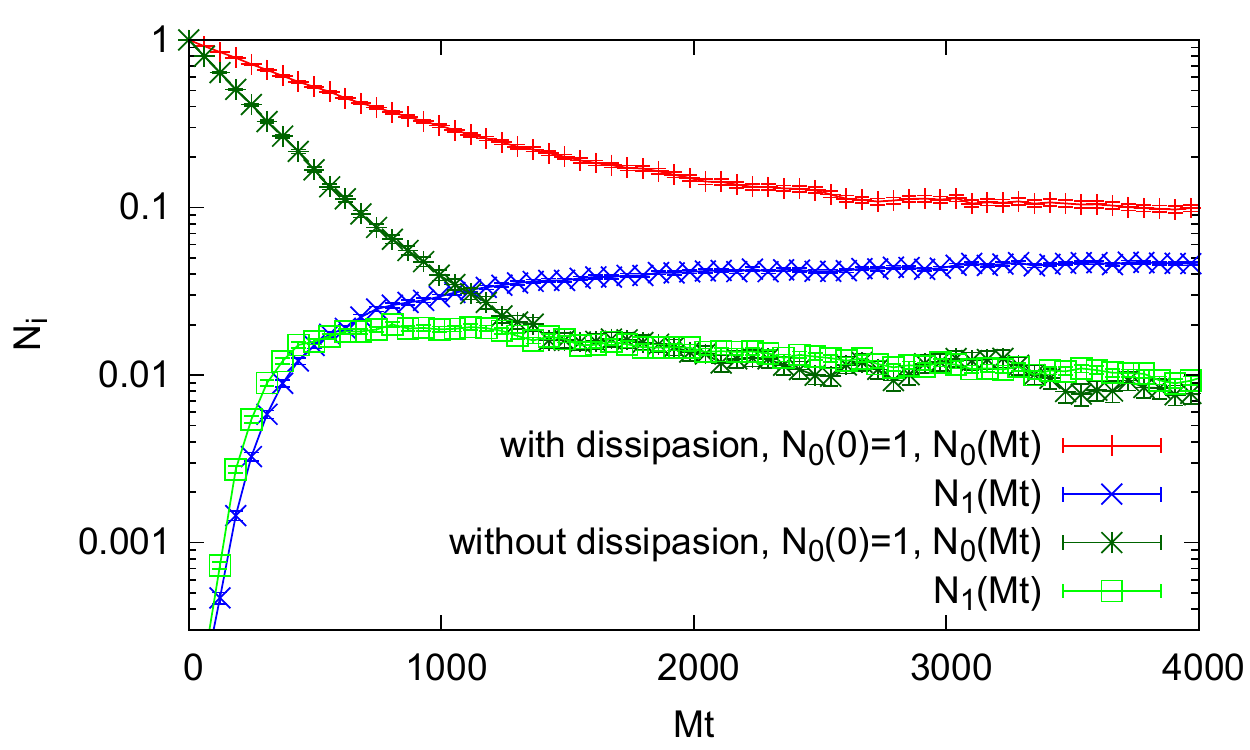}
\end{center}
\caption{
Effect of the dissipation on the occupation numbers. 
Shown is the time evolution of the occupation numbers of the ground
 state and the 1st excited state with and without dissipation. 
The bars represent statistical errors.
}
\label{fig:DEBYE_nodiss}
\end{figure}

As we have seen above, the relative motion is equilibrated with quantum dissipation.
To see the importance of the quantum dissipation, we switch off the $\mathcal O(T/M)$ terms, i.e. keeping only those terms explicit in \eqref{eq:qsd_intdiff}, and compare with the full simulation.
The comparison is made at $T/M=0.1$ and is shown in Fig.~\ref{fig:DEBYE_nodiss}.
We can see the clear differences not only in the asymptotic behavior on long time scales but also in the initial behavior.
When the dissipation is switched off, the initial decay of the ground state occupation is faster and approaches a much smaller value than in the case with dissipation.
Physically, the drag force prevents the heavy quark pair from dissociating in the QGP and balances with the thermal fluctuations to maintain the system in equilibrium.
Since we observe clear dissipative effects in the initial decay, we conclude that it may not be ignored even within the finite QGP lifetime $\sim 10$ fm. 
We will come back to this issue in a slightly more realistic setup in Sec.~\ref{sec:results_pheno}.

\subsection{Dependence of the temperature and the heavy quark mass}
\label{sec:results_tm}
\begin{figure}[htbp]
\begin{center}
\includegraphics[scale=0.8]{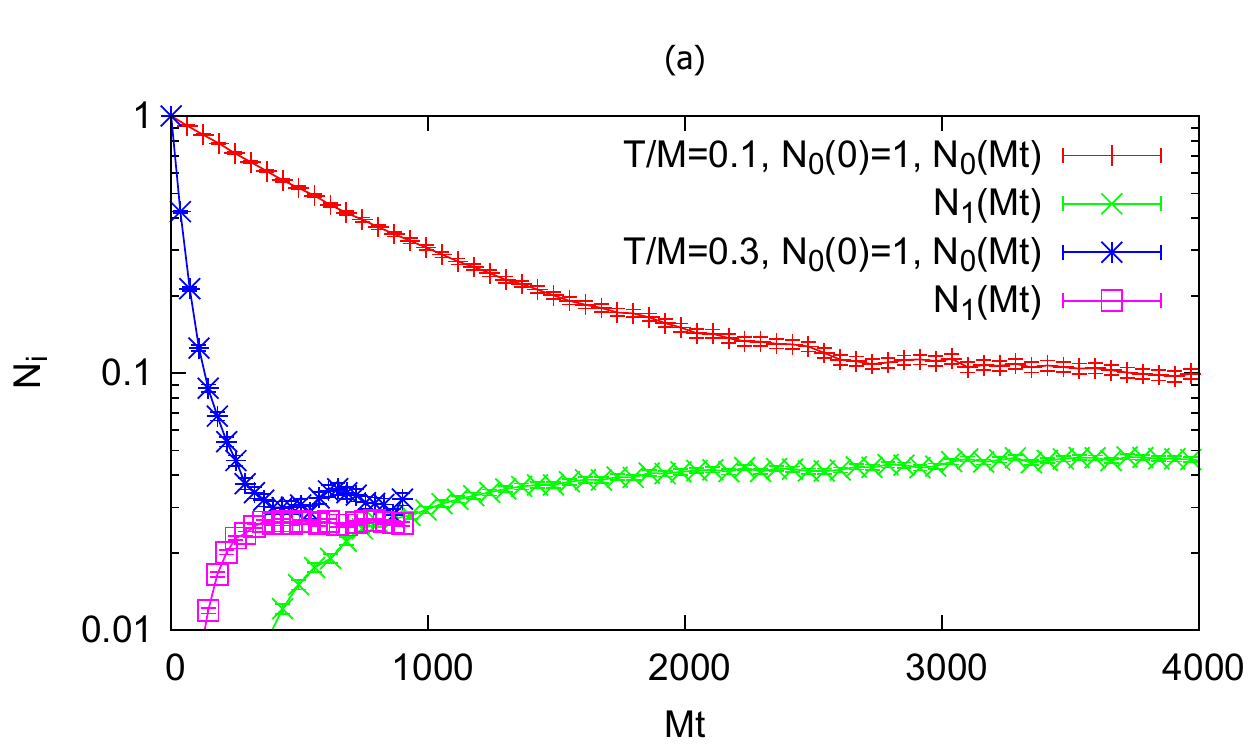}
\includegraphics[scale=0.8]{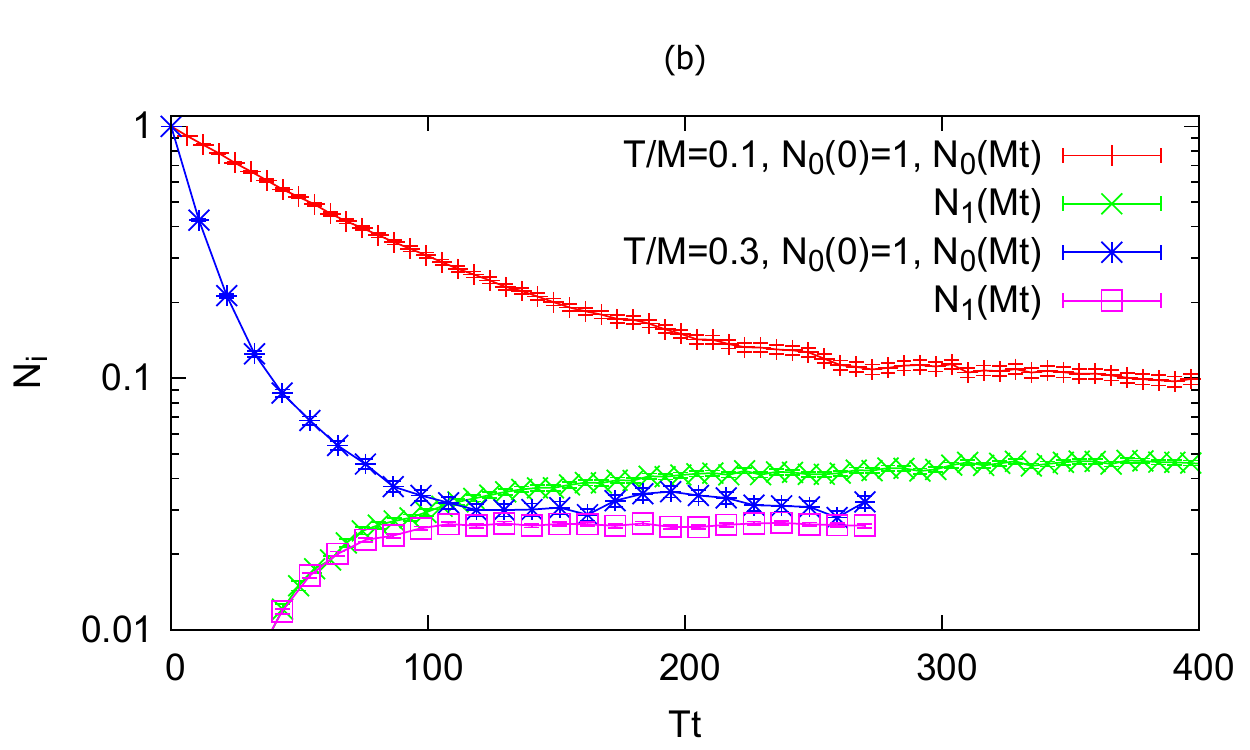}
\includegraphics[scale=0.8]{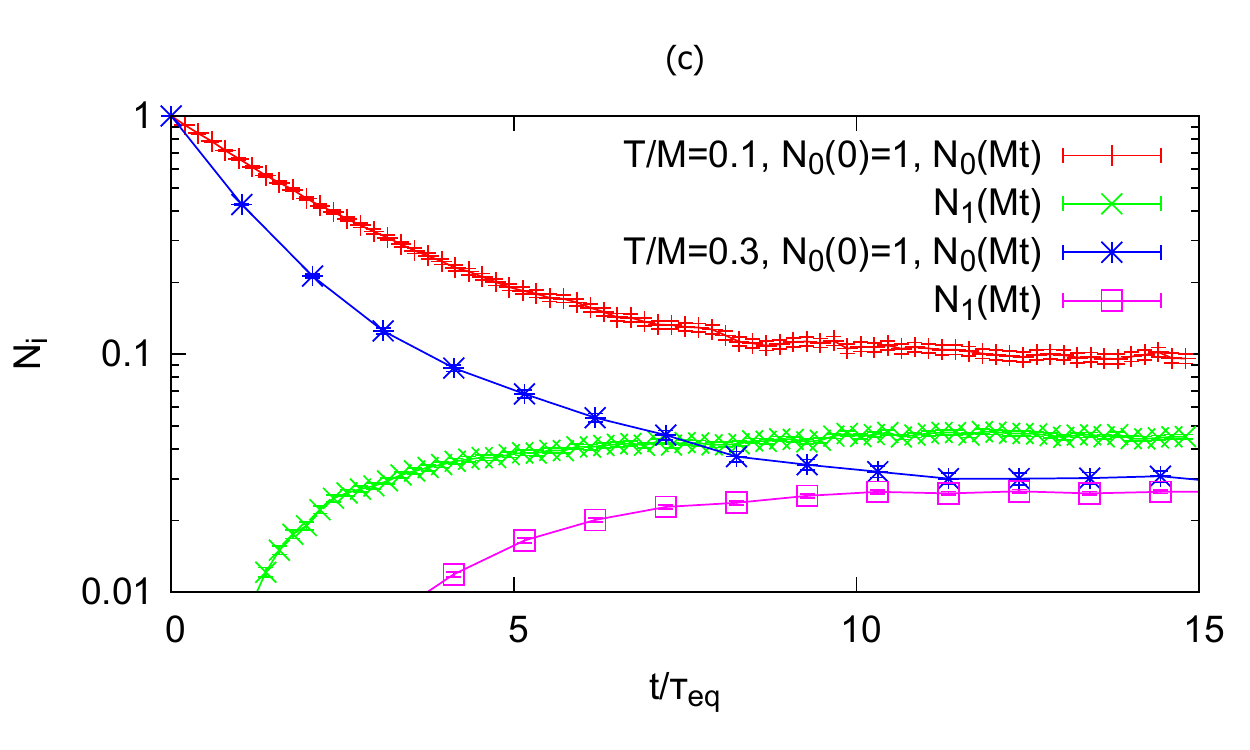}
\end{center}
\caption{
(a) Time evolution of the occupation numbers of the lowest 2 levels for $T/M=0.1$ and 0.3. 
In (b) and (c), time is rescaled by (b) the temperature $Tt$ and (c) the heavy quark damping rate $t/\tau_{\rm{eq}}$.
In (a) and (b), we show the results for $T/M=0.3$ only until the system reaches the steady state.
The bars represent the statistical errors. 
}
\label{fig:DEBYE_temperature}
\end{figure}

Here we study how the time evolution of the occupation numbers depends on the temperature.
We compare the results with $T/M=0.1$ and 0.3 starting from the ground state of $H_{\rm Debye}$ in Fig.~\ref{fig:DEBYE_temperature} (a).
For bottomonium, this corresponds to comparing $T\simeq 0.47$ GeV and
1.41 GeV, respectively.
As we can see in the figure, the relaxation takes place much faster at higher temperature $T/M=0.3$ because of two physical effects:  (i) The heavy quark damping rate is larger, and (ii) the ground state wave function is more extended  and receives decoherence more easily.
To cancel out the first effect, we plot the time scale in the unit of heavy quark damping rate $\tau_{\rm{eq}}\equiv MT \ell_{\rm{corr}}^2/\gamma =\pi (M/T)^2/M$ in Fig.~\ref{fig:DEBYE_temperature} (c).
There is still some difference between $T/M=0.1$ and 0.3, which we ascribe to the other effect: The decoherence of the ground state wave function.
The decoherence rate for a wave function of size $\ell_{\psi}\equiv
\sqrt{\langle x^2\rangle_{\psi}}$ is estimated as $\tau_{\rm
dec}^{-1}=D(0) - D(\ell_{\psi})$, which amounts to $\tau_{\rm dec}\simeq
456/M$ and $19/M$ for the ground state at $T/M=0.1$ and 0.3, respectively.
By rescaling to $\tau_{\rm eq}$, we get $\tau_{\rm dec}/\tau_{\rm
eq}=1.45$ and 0.55 for $T/M=0.1$ and 0.3, respectively, which qualitatively explains the reason why the initial decay for $T/M=0.3$ is faster even after rescaling to $\tau_{\rm eq}$\footnote{
To be strict, the inclusion of the dissipation changes the initial decay rate from the estimate by the decoherence rate as we saw in the previous section.
}. 

We can also interpret $T/M=0.1$ and 0.3 as a bottomonium and a
charmonium at $T=0.47$ GeV, respectively.
As shown in Fig.~\ref{fig:DEBYE_temperature} (b), we find the relaxation of a bottomonium proceeds more slowly than that of a charmonium again with the same two physical effects as above.

\subsection{Phenomenological implication to heavy ion collision experiments}
\label{sec:results_pheno}
In order to relate these simulations to quarkonia in heavy ion collisions, we take account of the expansion of the QGP by solving the QSD equation in a time-dependent environment undergoing Bjorken expansion:
\begin{align}
T(t)=T_0\Bigr(\frac{t_0}{t_0+t}\Bigr)^{\frac{1}{3}}, \quad
T_0=0.47 \ {\rm GeV}, \quad t_0 = 0.84 \ {\rm fm}.
\end{align}

We define occupation numbers by projecting the wave functions onto the
eigenstates of the Hamiltonian with the vacuum Cornell potential,
\begin{align}
H_{\rm Cornell} &= \frac{p^2}{M} -\frac{\alpha}{\sqrt{x^2+x_{\rm{c}}^2}}+\sigma x, \\
\alpha&=0.3, \quad x_c=\frac{1}{M}, \quad \sigma = 1.12 \ {\rm GeV / fm}, \quad
M=\left\{\begin{aligned}
&4.7 \ {\rm GeV} \quad \text{(bottom)}, \\
&1.6 \ {\rm GeV} \quad \text{(charm)},
\end{aligned}\right. \nonumber
\end{align}
so that the quarkonium yield is related to these occupation numbers at kinetic freezeout.

\begin{figure}[htbp]
\begin{center}
\includegraphics[scale=1.0]{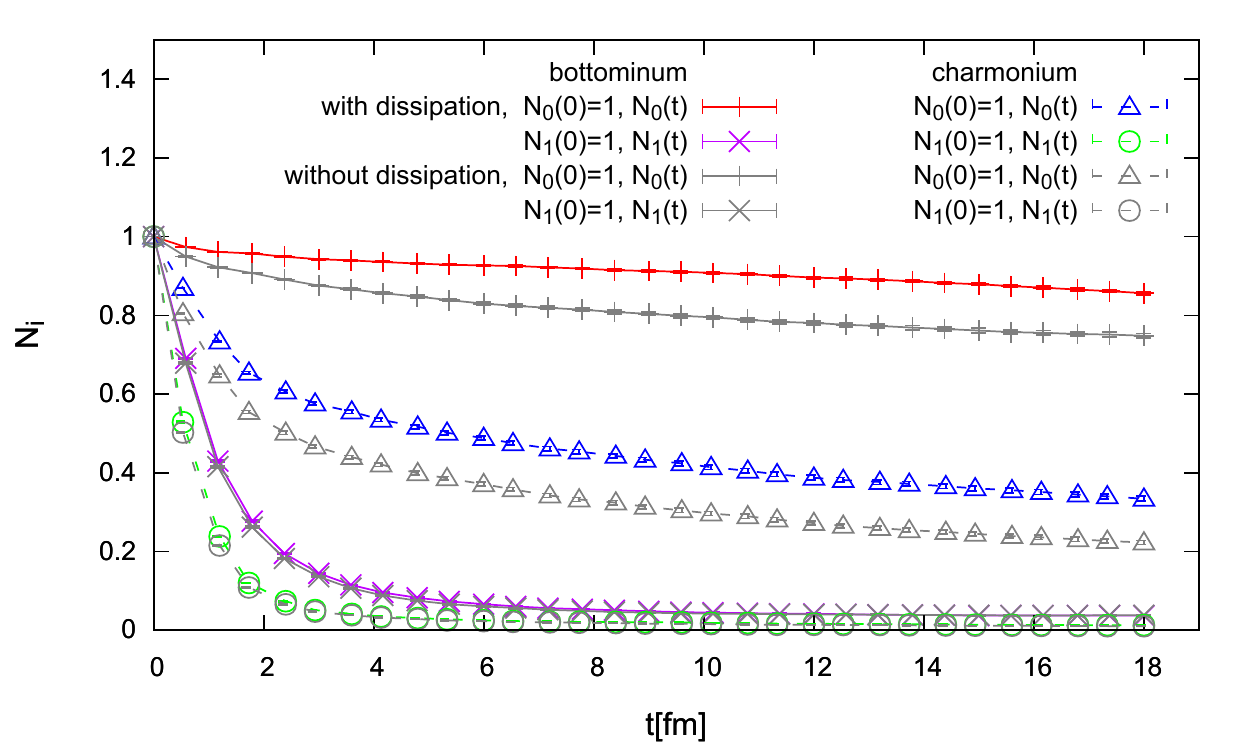}
\end{center}
\caption{
Time evolution of the occupation numbers of the ground state and the 1st excited state in Cornell potential.
The parameter is chosen to simulate bottomonium and charmonium in the Bjorken expanding QGP with the initial temperature $T_0=0.47 \ {\rm GeV}$ and the initial time $t_0=0.84 \ {\rm fm}$.
To see the effect of dissipation, we also plot the simulation without dissipation. 
The bars represent statistical errors.
}
\label{fig:bjorken}
\end{figure}

Note that we use the Debye screened potential for $V(x)$ in the QSD evolution equation.
Since the initial quarkonium wave function is not understood well and still an open issue, we adopt the eigenstates of $H_{\rm Cornell}$ in this study.
We take the ground state and the 1st excited state as the initial conditions and show in Fig.~\ref{fig:bjorken} how the occupation numbers of these states evolve in time. 
From Fig.~\ref{fig:bjorken}, we find that for both bottomonium and charmonium the occupation number of the 1st excited state decays faster than that of the ground state, which supports the phenomenological idea of sequential melting.
Also the charmonium ground state/1st excited state decays faster than the bottomonium ground state/1st excited state.
It is expected because bottomonium is more localized so that the decoherence is more ineffective and because bottomonium is heavier and thus the relaxation time is longer.

To see the effect of quantum dissipation, we also simulate without the dissipative terms as we did in Sec.~\ref{sec:results_equilibration} and plot in Fig.~\ref{fig:bjorken}.
Since the temperature at around 18~fm is about the transition temperature 170~{MeV}, \footnote{
The typical QGP lifetime is $\sim 10~{\rm fm}$ in full three dimensional hydrodynamic simulations.
} the dissipative effects on the relative motion of a quarkonium in the QGP are up to 20\% effect for the ground states and marginally effective for the 1st excited state.
The reason why dissipation affects the ground state more is that the
decoherence is ineffective for localized states and that the relative importance of the dissipation enhances as we found in \cite{Akamatsu:2018xim}. 
By comparing with Fig.~\ref{fig:DEBYE_nodiss}, one may notice that the dissipative effects appear to be much less important here in Fig.~\ref{fig:bjorken}.
However, when making a comparison, one needs to consider the temperature decrease for the latter because the decoherence and damping proceed much slower at lower temperatures.

\section{Summary}
\label{sec:summary}
In this paper, we study the quantum Brownian motion of a heavy quark pair in the quark-gluon plasma (QGP).
Treating the heavy quark pair as an open quantum system in the QGP, we
derive and solve the master equation for the relative motion among the pair.
In this master equation, three different forces govern the evolution of the heavy quarks: Debye screened potential force, thermal noise, and drag force.
Thus the classical counterpart of our quantum mechanical description is given by the Langevin equation of two Brownian particles interacting with each other through the Debye screened potential.
It is for the first time that a full quantum mechanical simulation
including dissipation is performed for heavy quarks in the QGP.

The master equation of the heavy quark pair takes the Lindblad form \eqref{eq:Lindblad_rel}, which ensures the basic physical properties of the reduced density matrix: positivity, hermiticity, and unitarity.
Among the three different forces stated above, the Lindblad operator is responsible for the thermal noise and the drag force.
The thermal noise is represented by a momentum shift operator while the
drag force by the recoil of heavy quarks during each microscopic collision.
The Lindblad operator for the relative motion of the heavy quark pair depends on the center-of-mass momentum, but its dependence is found to be small so that we only considered the static case ($P_{\rm CM}=0$) in our simulations.

We solved the Lindblad master equation by a stochastic unravelling method called quantum state diffusion (QSD).
Using this method, any Lindblad master equation is shown to be equivalent to a nonlinear stochastic Schr\"odinger equation by which we can correctly produce a mixed state ensemble for the density matrix.
In our numerical simulation in one dimension, we first checked basic properties of the master equation (Sec.~\ref{sec:results_equilibration}).
The occupation number of eigenstates relaxes toward a value independent of the initial condition, and the steady state distribution is consistent with the Boltzmann distribution.
We also studied the effect of quantum dissipation by comparing a simulation without dissipation.
The quantum dissipation delays the relaxation toward equilibrium, which
is consistent with our intuitive classical picture that the drag force
prevents a heavy quark pair from dissociating.
We next simulated the temperature and heavy quark mass dependences on the time evolution (Sec.~\ref{sec:results_tm}).
The time evolution strongly depends on these quantities but is found to
scale to a considerable extent with the heavy quark damping time.
Finally, we take into account the expansion of the QGP in relativistic heavy ion collisions and solved the master equation with a time-dependent temperature (Sec.~\ref{sec:results_pheno}).
By simulating the bottomonium and charmonium yields in the QGP undergoing the Bjorken expansion, we found that charmonium dissociates faster than bottomonium and that the excited states decay faster than the ground states.
We also found that the quantum dissipation delays the ground state dissociation compared to the case without dissipation while the excited state dissociation is totally insensitive to the quantum dissipation.
This difference comes from the fact that for an extended state the decoherence due to thermal fluctuation is the driving force of the dissociation while for a localized state (such as the ground state) the decoherence is ineffective and as a result the dissipation becomes as important as the decoherence.

Through this analysis, it becomes clear that quarkonia probe fundamental length scales of the QGP: the screening length $1/m_{\rm D}$ and the correlation length $\ell_{\rm corr}$ as shown in \eqref{eq:VD_parametrization}.
The screening length has been evaluated via the real-part of the heavy quark potential (see e.g. \cite{burnier2016gauge,Burnier:2015tda,Lafferty:2019jpr}) and the heavy quark free energies (see e.g. \cite{Maezawa:2011aa,maezawa2012application}) in various lattice QCD simulations. On the other hand, the correlation length is related to the imaginary part, whose determination from the lattice is much less robust.
Using heavy quark observables in heavy-ion collisions such as single leptons and open heavy flavor mesons, the single heavy quark damping rate has been determined phenomenologically with some accuracy. The corresponding quantity in the quarkonium Lindblad equation is given by $\tau_{\rm eq}=MT\ell_{\rm corr}^2/\gamma$.
Therefore, the suppression pattern of quarkonium yields provides a way to determine the correlation length of colored excitations $\ell_{\rm corr}$, a fundamental dynamical quantity of the QGP which has not yet been calculated precisely on the lattice.
To perform such a comprehensive study in the future, we need to implement a more realistic computation, such as a three-dimensional simulation on a three-dimensional hydrodynamic background and include the color of the heavy quarks, to name a few.

\begin{acknowledgments}
T.M. is grateful to Shiori Kajimoto for valuable discussions.
The work of Y.A. is supported by JSPS KAKENHI Grant Number JP18K13538.
M.A. is supported in part by JSPS KAKENHI Grant Number JP18K03646.
This work has utilized computing resources provided by  
UNINETT Sigma2 - the National Infrastructure for High Performance Computing and Data Storage in Norway under project NN9578K-QCDrtX "Real-time dynamics of nuclear matter under extreme conditions"
\end{acknowledgments}

\appendix
\section{Dependence of the center-of-mass motion}
\label{sec:results_com}
\begin{figure}[htbp]
\begin{center}
\includegraphics[scale=1.0]{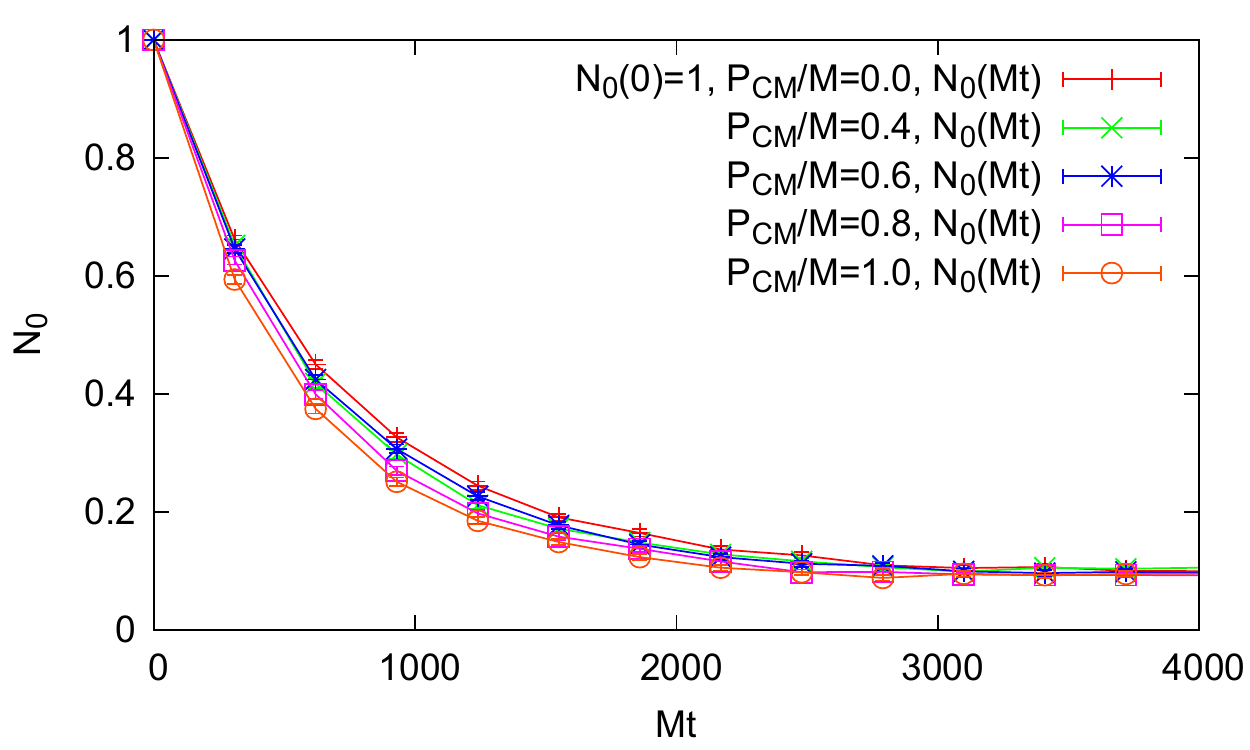}
\end{center}
\caption{
Effects of the center-of-mass motion on the ground state occupation number.
The center-of-mass momenta are $P_{\rm{CM}}/M=0, 0.4, 0.6, 0.8$, and $1.0$.
The bars represent the statistical errors.
}
\label{fig:DEBYE_PCM}
\end{figure}

\begin{figure}[htbp]
\begin{center}
\includegraphics[scale=1.0]{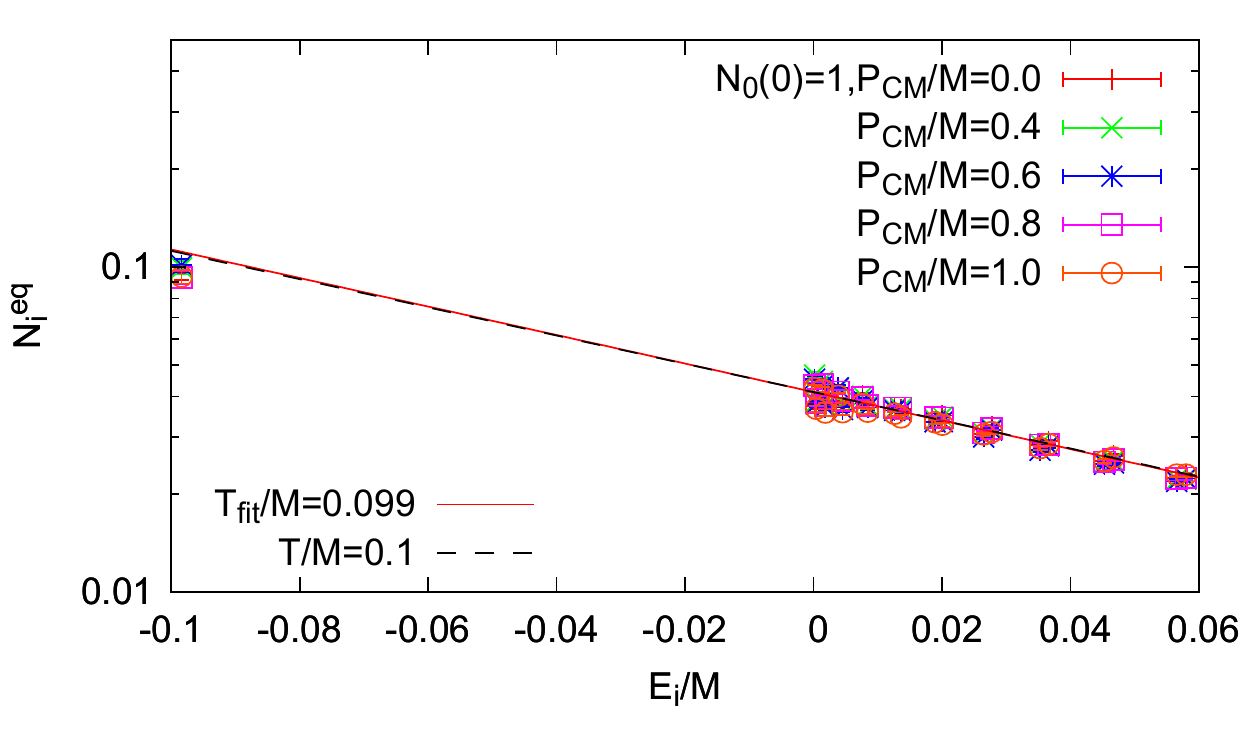}
\end{center}
\caption{
Effects of the center-of-mass momentum on the eigenstate occupation numbers at late enough time $Mt=4650$.
The center-of-mass momenta are $P_{\rm{CM}}/M=0, 0.4, 0.6, 0.8$, and $1.0$.  
The bars represent statistical errors.
}
\label{fig:main_DEBYE_PCM_eq}
\end{figure}

Here we show the effects of the center-of-mass motion on the relative motion of a quarkonium in the QGP at $T/M=0.1$.
We set the center-of-mass momentum at different values $P_{\rm{CM}}/M=0, 0.4, 0.6, 0.8$, and
$1.0$, which are equivalent to the center-of-mass velocity
$v_{\rm{CM}}=0, 0.2, 0.3$, $0.4$, and $0.5$, respectively. 
The time evolution for the ground state occupation numbers is shown
in Fig.~\ref{fig:DEBYE_PCM}.
The result shows that the effect of $P_{\rm CM}$ is rather small.
In Fig.~\ref{fig:main_DEBYE_PCM_eq}, we also show the occupation number
for each eigenstate at late enough time ($Mt=4650$) when the system has
reached a (non-equilibrium) steady state.
Again, the result is quite insensitive to the value of $P_{\rm CM}$.

\section{Explicit form of QSD equation}
\label{sec:qsdeqform}
We present here the explicit form of the QSD equation \eqref{eq:qsd2} with heavy quark colors ignored:
\begin{align}
d\psi(\vec{x})
=&\hspace{0.3cm}
dt\Bigr[
i\frac{\vec{\nabla}^2}{M}\psi(\vec{x})-iV(\vec{x})\psi(\vec{x})
\Bigr]
\notag\\
&+\frac{2dt}{\bra{\psi}\psi\rangle}\Bigr\{
\notag
\int d\vec{y}\,\,\Bigr(
\tilde{G}_1(\vec{x},\vec{y})
N(\vec{y})
\psi(\vec{x})
+\vec{\tilde{H}}_1(\vec{x},\vec{y})
\cdot N(\vec{y})
\,\,\vec{\nabla}_x\psi(\vec{x})
\notag\\
&\hspace{3cm}
+\vec{\tilde{G}}_2(\vec{x},\vec{y})
\cdot
\vec{J}(\vec{y})\,\,
\psi(\vec{x})
+\tilde{H}_2^{ij}(\vec{x},\vec{y})\,
{J}^j(\vec{y})\,
\nabla^i_x\psi(\vec{x})\Bigr)
\Bigr\}
\notag\\
&-dt\Bigr[
\hspace{0.15cm}
I_1(\vec{x})\psi(\vec{x})+\vec{I}_2(\vec{x})\cdot\vec{\nabla}\psi(\vec{x})
+I_3^{ij}(\vec{x})\nabla^i\nabla^j\psi(\vec{x})\Bigr]
\notag\\
&+\Bigr[
\zeta_1\Bigr(\frac{\vec{x}}{2}\Bigr)\psi(\vec{x})-\vec{\zeta}_2\Bigr(\frac{\vec{x}}{2}\Bigr)\cdot\vec{\nabla}\psi(\vec{x})
\Bigr]\,,
\end{align}
with
\begin{align}
G_1(\vec{x})&=D(\vec{x})+\frac{\vec{\nabla}^2}{8MT}D(\vec{x})\,,
\\
\vec{G}_2(\vec{x})&=\vec{\nabla}D(\vec{x})+\frac{\vec{\nabla}\vec{\nabla}^2}{8MT}D(\vec{x})\,,
\\
\tilde{G}_1(\vec{x},\vec{y})&=
G_1\Bigr(\frac{\vec{x}-\vec{y}}{2}\Bigr)-G_1\Bigr(\frac{\vec{x}+\vec{y}}{2}\Bigr)\,,
\\
\vec{\tilde{G}}_2(\vec{x},\vec{y})&=
\frac{1}{4MT}\Bigr[
\vec{G}_2 \Bigr(\frac{\vec{x}-\vec{y}}{2}\Bigr)+\vec{G}_2
\Bigr(\frac{\vec{x}+\vec{y}}{2} \Bigr)
\Bigr]\,,
\\
\vec{\tilde{H}}_1(\vec{x},\vec{y})&=
\frac{1}{4MT}\Bigr[
[\vec{\nabla}D]\Bigr(\frac{\vec{x}-\vec{y}}{2}\Bigr)-[\vec{\nabla}D]\Bigr(\frac{\vec{x}+\vec{y}}{2}\Bigr)
\Bigr]\,,
\\
\tilde{H}_2^{ij}(\vec{x},\vec{y})&=
\frac{1}{16M^2T^2}
\Bigr[
[\nabla^i\nabla^jD]\Bigr(\frac{\vec{x}-\vec{y}}{2}\Bigr)
+
[\nabla^i\nabla^jD]\Bigr(\frac{\vec{x}+\vec{y}}{2}\Bigr)
\Bigr]\,,
\\
I_1(\vec{x})&=D(\vec{0})-D(\vec{x})
+\frac{\vec{\nabla}^2D(\vec{0})}{4MT}
+\frac{(\vec{\nabla}^2)^2D(\vec{0})}{64M^2T^2}
+\frac{(\vec{\nabla}^2)^2D(\vec{x})}{64M^2T^2}
+\frac{\vec{\nabla}^2D(\vec{x})}{4MT}\,,
\\
\vec{I}_2(\vec{x})&=
\frac{\vec{\nabla}\vec{\nabla}^2D(\vec{x})}{16M^2T^2}-\frac{\vec{\nabla}D(\vec{x})}{2MT}\,,
\\
I_3^{ij}(\vec{x})&=
\frac{\nabla^i\nabla^jD(\vec{0})}{16M^2T^2}
+\frac{\nabla^i\nabla^jD(\vec{x})}{16M^2T^2}\,,
\\
N(\vec{x})&=\psi^{\dagger}(\vec{x})\psi(\vec{x})\,,
\\
\vec{J}(\vec{x})&=\psi^{\dagger}(\vec{x}){\vec{\nabla}}\psi(\vec{x})\,,
\\
\zeta_1(\vec{x})&=d\xi^{}(\vec{x})-d\xi^{}(-\vec{x})+ \frac{\vec{\nabla}^2d\xi^{}(\vec{x})}{2MT}-
\frac{\vec{\nabla}^2d\xi^{}(-\vec{x})}{2MT}\,,
\\
\vec{\zeta}_2(\vec{x})&=\frac{\vec{\nabla}d\xi^{}(\vec{x})}{2MT}-
\frac{\vec{\nabla}d\xi^{}(-\vec{x})}{2MT}\,.
\end{align}

\section{Equilibrium distribution in the classical limit}
\label{sec:eqdist_classical}
The explicit form of the quantum master equation for a heavy quark and antiquark pair has been given in \cite{Akamatsu:2014qsa}.
By considering up to the 2nd order in the derivative expansion for the time coarse graining, we obtain the influence functional $S_{\rm IF} = S_{\rm pot} + S_{\rm fluct} + S_{\rm diss} + S_{\rm L}$, where each term corresponds to a specific power counting in the heavy quark velocity $v$ as $S_{\rm pot}\sim S_{\rm fluct}\sim v^0$, $S_{\rm diss}\sim v$, and $S_{\rm L}\sim v^2$.
Let us ignore $S_{\rm L}$, which is justified when the velocity is small $v\ll 1$, and take the classical limit in the master equation.
For simplicity, we do not consider color degrees of freedom here.
Then, in contrast to the fully quantum system considered in the main text, the master equation for the center-of-mass motion can be integrated.
The result is
\begin{align}
\label{eq:master1}
\partial_t \rho(x,y)
=&\left[
\frac{i}{M}\left(\vec{\nabla}_x^2 - \vec{\nabla}_y^2 \right)
- i\left\{V(x) - V(y)\right\}
\right]\rho(x,y) \\
&+\left[
2F_1\left(\frac{x-y}{2}\right) -2F_1(0)
+F_1(x) + F_1(y) - 2F_1\left(\frac{x+y}{2}\right)
\right]\rho(x,y) \nonumber \\
&+\left[
\begin{aligned}
&2\vec F_2\left(\frac{x-y}{2}\right)\cdot\left(\vec\nabla_x - \vec \nabla_y\right)
+2\vec F_2(x)\cdot \vec\nabla_x + 2\vec F_2(y)\cdot \vec\nabla_y \\
&-2\vec F_2\left(\frac{x+y}{2}\right)\cdot\left(\vec\nabla_x + \vec \nabla_y\right)
\end{aligned}
\right]\rho(x,y), \nonumber
\end{align}
with
\begin{align}
\label{eq:f12}
F_1(x) \equiv D(x) + \frac{\vec{\nabla}^2 D(x)}{4MT}\simeq D(x), \quad
\vec F_2(x) \equiv \frac{\vec\nabla D(x)}{4MT}\simeq \frac{\vec \nabla F_1(x)}{4MT}.
\end{align}
Using the coordinates
\begin{align}
R = \frac{x+y}{2}, \quad
r = x-y,
\end{align}
it is written as
\begin{align}
\label{eq:master2}
\partial_t \rho(R,r)
=&\left[
\frac{2i}{M}\vec\nabla_R\cdot\vec\nabla_r
- i\left\{V\left(R+\frac{r}{2}\right) - V\left(R-\frac{r}{2}\right)\right\}
\right]\rho(R,r) \\
&+\left[
2F_1\left(\frac{r}{2}\right) -2F_1(0)
+F_1\left(R+\frac{r}{2}\right) + F_1\left(R-\frac{r}{2}\right) - 2F_1(R)
\right]\rho(R,r) \nonumber \\
&+\left[
\begin{aligned}
&\left\{4\vec F_2\left(\frac{r}{2}\right) + 2\vec F_2\left(R+\frac{r}{2}\right) - 2\vec F_2\left(R-\frac{r}{2}\right)\right\}\cdot\vec\nabla_r \\
&+\left\{\vec F_2\left(R+\frac{r}{2}\right) + \vec F_2\left(R-\frac{r}{2}\right) - 2\vec F_2(R)\right\}\cdot\vec\nabla_R
\end{aligned}
\right]\rho(R,r). \nonumber
\end{align}

In the classical limit, we can prove that the equilibrium distribution is the Boltzmann distribution.
By performing the Wigner transformation
\begin{align}
f_p(R) = \int d^3 r \exp\left[-i\frac{\vec p\cdot \vec r}{\hbar}\right] \rho(R,r),
\end{align}
and taking the classical limit $\hbar\to 0$
\footnote{
When the dimension of $\hbar$ is recovered, the kinetic term becomes $\nabla_R\nabla_r \hbar$, potential $V/\hbar$, fluctuation $F_1/\hbar^2$, and dissipation $F_2$ is unchanged.
}, we obtain the classical kinetic equation:
\begin{align}
\partial_t f_p(R) = 
\left[
\begin{aligned}
&-\frac{2}{M}\vec p\cdot\vec \nabla_R + \vec \nabla V(R) \cdot \vec\nabla_p \\
&-\frac{1}{4}\left\{\partial_i\partial_j F_1(0) + \partial_i\partial_j F_1(R)\right\}\partial^p_i\partial^p_j \\
&-2\left\{\partial_i F_{2j}(0) + \partial_i F_{2j}(R)\right\}\partial^p_i p_j
\end{aligned}
\right]f_p(R).
\end{align}
Using the approximation $\vec F_2(x)=\vec\nabla F_1(x)/4MT$ employed in \eqref{eq:f12}, the equilibrium distribution is
\begin{align}
f_p^{\rm eq}(R) \propto \exp\left[
-\frac{1}{T}\left\{\frac{p^2}{M} + V(R)\right\}
\right].
\end{align}
If we do not use this approximation, the correction is estimated as
\begin{align}
F_1(x)\sim D(x) \left(1 - \frac{1}{4MT\ell_{\rm corr}^2}\right) \sim D(x)\left(1 - \frac{T}{4M} \right),
\end{align}
and the effective temperature slightly deviates from the environment temperature,
\begin{align}
T_{\rm eff} \equiv \frac{\vec\nabla F_1(x)/4M}{\vec F_2(x)} \sim T\left(1-\frac{T}{4M}\right),
\end{align}
which gives a rough explanation of the difference between $T_{\rm fit}$ and $T$ in Fig.~\ref{fig:main_DEBYE_temperature}.

\nocite{*}

\bibliography{final3_PRD}
\end{document}